\documentclass{pasj00}
\usepackage{txfonts}

\def\aj{\rm{AJ}}                    
\def\apj{\rm{ApJ}}                 
\def\apjl{\rm{ApJ}}                
\def\apjs{\rm{ApJS}}                        
\def\mnras{\rm{MNRAS}}   

\usepackage{epsfig}
\usepackage{natbib}
\bibpunct{(}{)}{;}{a}{}{,} 

\begin{document}




\author{Charles L. \textsc{Steinhardt}}
\affil{Kavli IPMU, University of Tokyo, Kashiwanoha 5-1-5, Kashiwa, Chiba, Japan 277-8583}
\email{charles.steinhardt@ipmu.jp}

\author{John D. \textsc{Silverman}}
\affil{Kavli IPMU, University of Tokyo, Kashiwanoha 5-1-5, Kashiwa, Chiba, Japan 277-8583}
\email{john.silverman@ipmu.jp}

\title{Quasars with Anomalous H$\beta$ Profiles I: Demographics}





\maketitle


\begin{abstract}
The H$\beta$ emission line in a typical Type I quasar is composed of a broad base and a narrow core, with the core velocity typical of narrow-line region emission, and line-fitting routines typically assume this picture.  We test the effects of removing this constraint, and find a substantial group of Type I quasars in the Sloan Digital Sky Survey catalog with H$\beta$ emission line cores broader than 1200 km/s, above the velocity believed possible for gas in the quasar narrow-line region.   We identify this group of ``anomalous H$\beta$ quasars'' (AHQs) as a distinct population because of a variety of spectral and photometric signatures common to these AHQs but atypical of other quasars.  These features are similar to some aspects of narrow-line Seyfert 1s and correlations identified by Eigenvector 1, but also contain distinct features that make AHQs difficult to classify.  We demonstrate that AHQs comprise at least 11\% and most likely approximately one quarter of the SDSS Type I quasar population at $0.2 < z < 0.8$.  For AHQs, the [O{\small III}]$\lambda 4959,5007$ profile is often better fit by de-linking it from the H$\beta$ core, while a more standard linked fit produces a tight correlation between narrow- and broad-line velocities.   We find that [O{\small III}] in AHQs sometimes has a standard narrow-line profile and other times matches the H$\beta$ core, but is rarely in between the two, implying that the broadened core emission arises from a distinct physical region.  Another feature of AHQs is a diminished [O{\small II}] line, which might indicate a connection between AHQs and the interstellar mediums of their host galaxies, through reduced photoionization or star formation.  We find that it is difficult to produce AHQs using the current quasar standard model.
\end{abstract}


\section{Introduction}

Although quasars are luminous enough to be detected at large distances, they are too small to be resolved into multiple spatial elements except at very short distances (e.g., \citet{Marconi2006}) or when lensed by a foreground object \citep{Sluse2011}.  As a result, our best tool for mapping the region around the central supermassive black hole in higher-redshift active galactic nuclei has been spectroscopy.  Each spectral line is associated with a specific ionization potential, and the shape of the line corresponds to the peculiar velocity profile of the emitting gas, complicated by parameters such as densities, abundances, temperatures, and effects of radiative transfer.  There is a long history of studying quasar emission lines and their correlations (cf. \citet{Boroson1992}), much of which has helped to develop our standard model for Type I quasars  \citep{Sulentic2000, Gaskell2009}.  In this paper, we describe a new class of quasars with distinct spectroscopic features that do not appear to fit easily into that standard model.

The standard picture built from existing data consists, from the black hole outward, of (cf. \citet{Petersonbook}):
\begin{itemize}
\item{The central black hole}
\item{A hot accretion disk, beginning near the "innermost stable circular orbit" (ISCO), typically a few Schwarzchild radii away with an exact radius dependent upon the spin of the central black hole.  The accretion disk and corona are believed to emit the quasar continuum.}
\item{The "broad-line region" (BLR), $\sim 0.1-1$ pc away, from which high-velocity gas produces correspondingly broad spectral emission lines, with typical FWHM in the 2000-20000 km/s range.  Prominent emission lines visible in optical spectra at $z > 0.5$ include, from ionization potential placing them nearest to the central black hole, C{\small IV}, a broad component of H$\beta$, and Mg{\small II}.  In quasar virial mass estimators \citep{McLure2002,McLure2004,Vestergaard2006,Wang2009,Onken2008,Risaliti2009,Rafiee2011}, the velocities of BLR gas are assumed to be predominantly virial in order to use Kepler's Laws to infer the mass of the central black hole.  However, C{\small IV} in particular may also be substantially broadened by radiation pressure and quasar outflows \citep{Marconi2009,Netzer2010,Marziani2011,Netzer2009}.}
\item{The "narrow-line region" (NLR), $\sim$ kpc away, from which lower-velocity gas produces correspondingly narrower spectral emission lines, with typical FWHM approximately 500 km/s,, and in some cases higher velocities due to thermal or other effects.  This FWHM might correspond to typical velocities for gas in the interstellar medium being photoionized by the central black hole.  Prominent narrow lines in the optical include the [O{\small III}] fine structure doublet, a narrow component of H$\beta$, and usually [O{\small II}].  }
\item{A surrounding host galaxy.  For quasars, the galactic starlight is typically too faint compared to the quasar to be detected directly, but galactic spectral lines are often present, including both absorption lines and, for many galaxies, a narrow [O{\small II}] emission line believed to be associated with star formation.  In a typical quasar spectrum, the [O{\small II}] emission is likely dominated by photoionization due to a non-thermal continuum as opposed to star formation.}
\end{itemize}

For some individual active galactic nuclei, reverberation mapping \citep{Bentz2009,Peterson2004} has been able to confirm the inner portion of this picture, out to the H$\beta$ broad emission line.  In a time series of spectra for the same object, a increase in the continuum luminosity is followed, often hundreds of days later, by a similar flare in C{\small IV} and then H$\beta$.  Assuming the flare propagates outward at the speed of light, the delay can be used to infer a radius to BLR spectral lines.  

It should be noted, however, that this simple picture is merely a broad overview of features common to most quasars.  Most individual quasars are observed to deviate from this model in any of a wide variety of ways, particularly where high-quality spectra are available and particularly outside of the broad-line region.

There have also been hints that for some quasars, this simple picture might be more clearly wrong.  The \citet{Boroson1992} principal component analysis uncovered Eigenvector 1, indicating that for some quasars, changes in specific broad spectral lines are correlated with changes in specific narrow lines.  The population of narrow-line Seyfert 1 galaxies contain active nuclei with no traditional broad-line component, but with an intermediate-velocity component up to $\sim 2000$ km/s \citep{Komossa2008}.  It is unclear whether this intermediate component is a broadened narrow-line region, a weak, high-radius broad-line region, or something else entirely.  Similarly, some quasars contain an intermediate H$\beta$ component \citep{Hu2008}.  Even the [O{\small III}] narrow line ($\lambda 4959, 5007$) has been observed to occasionally be broader than 1000 km/s \citep{Marziani1996,McIntosh1999}.  

The [O{\small III}] fine structure doublet is an especially useful indicator for properties of Seyfert galaxies and other low-luminosity active galactic nuclei (AGNs) \citep{Hao2005}.  [O{\small III}] is a forbidden line that can only be produced in low-density gas where there are not enough free electrons to allow competing transitions.  As such, [O{\small III}] lines are typically found around AGN only in the narrow line region, as interstellar medium gas is lower-density than gas in the broad-line region.  Common line-fitting techniques for the nearby H$\beta$ line rely upon the assumption that the widths of the narrow H$\beta$ component and [O{\small III}] lines are identical, in order to disentangle the complicated combination of spectral lines surrounding H$\beta$.  The [O{\small III}] lines, in turn, are typically assumed to be narrow based upon previous studies \citep{Hao2005}.  For example, the \citet{Shen2008} virial mass catalog fit the H$\beta$ broad component along with a narrow component linked to the [O{\small III}] line width and at a maximum FWHM of $1200$ km/s.

Although fitting each of these narrow lines as one Gaussian, with a common width, is currently a standard practice, there is ample evidence that the underlying physics may be more complicated.  Forbidden lines at higher ionization potentials do show broader velocity profiles then those with lower ionization potentials \citep{Whittle1985,DeRobertis1984}, indicating that the narrow-line region may not be monolithic.  [O{\small III}], a narrow line heavily studied due because it is typically strong and available in the optical up to $z \sim 0.8$, may be asymmetric with multiple components \citep{Zamanov2002,Boroson2005}.  This might be further evidence of an intermediate-line region, or of outflows (cf. \citet{Hall2002,Ho2009}).
  
Motivated by these examples of broader H$\beta$ cores and non-Gaussian narrow-lines, in this work we examine the effects of removing the 1200 km/s bound from \citet{Shen2008}.  {\em If the H$\beta$ line is indeed composed predominantly of one broad and one narrow component as expected theoretically, removing this bound should have no substantial effect.}  Shen et al. (2011) model complex line shapes by adding a third and even fourth Gaussian in an attempt to describe the H$\beta$ profile.  Instead, we find and describe a set of quasars for which the $\chi^2/\textrm{DOF}$ is substantially lowered when fit with a broadened ``narrow" component of H$\beta$ (\S~\ref{sec:sample}), but also that these two components are still sufficient to describe the entire H$\beta$ profile; a third component is not necessary. 

 It turns out these ``anomalous H$\beta$ quasars'' (AHQs, see \S~\ref{sec:properties}) comprise a distinct population, with a wide variety of properties common to all AHQs but atypical of other quasars.  These properties are discussed in \S~\ref{sec:properties}, and include many of the characteristics previously reported by \citet{Boroson1992} and \citet{Hu2008}, as well as several properties characteristic of narrow-line Seyfert 1s, despite our study focusing on Type I quasars often containing a 10000 km/s broad H$\beta$ component.  In addition, we show that AHQ spectra have several additional, previously unreported properties characteristic of this new population.  We consider whether these properties might be an artifact of our fitting routines in \S~\ref{sec:fitting}.  Finally, AHQs seem not to fit easily into the standard AGN model, and possible explanations are considered in \S~\ref{sec:discussion}, as well as issues of nomenclature.  It is currently unclear how AHQs should be classified, but likely that a physical model will demonstrate how to fit them into existing quasar nomenclature.

\section{Anomalous H$\beta$ Profiles}
\label{sec:sample}

\begin{figure*}
  \vskip1.5cm
  \epsfxsize=7in\epsfbox{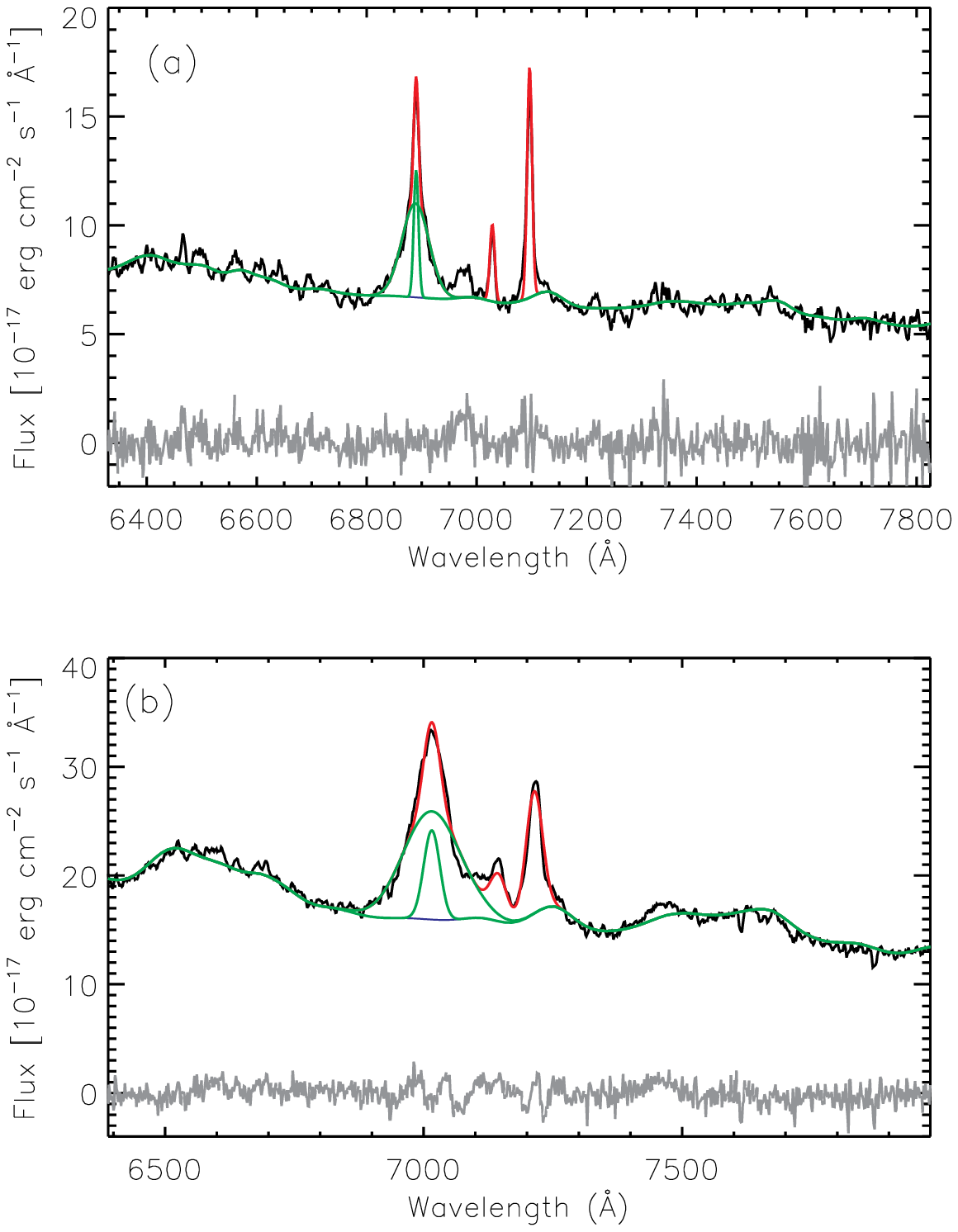} 
  \caption{Best-fit sum of continuum and iron (green) and total fit (red) to (a) SDSS J000109.12-004121.6, a standard, or `uncorrelated' quasar, and (b) SDSS J000545.61+153833.8, a `correlated' quasar with a broad H$\beta$ core.  Both H$\beta$ components are also shown.  The residual differences between spectrum (black) and fit are shown in gray for the original spectrum (the depicted spectra are boxcar-smoothed).}
  \label{fig:samples}
\end{figure*}

Fitting techniques currently in use typically assume that the H$\beta$ core (i.e., the narrower component) and [O{\small III}] both come from the narrow-line region, so that the two can be linked.  Methods for linking them include fitting the [O{\small III}] line and using it as a tempate for a narrow H$\beta$ component and fitting the two components simultaneously.  Because one of the main recent uses of quasar H$\beta$ emission has been as a virial mass indicator, we will begin our investigation using the technique on which the SDSS Value Added Catalog \citep{Shen2011,Shen2008} and calibration of Mg{\small II} virial masses against H$\beta$ \citep{McLure2004} are based: 
\begin{enumerate}
\item{Using the catalog redshift, we select the spectrum at rest wavelengths $[4435,4700]$ and $[5100,5535]$ \AA~and simultaneously fit the sum of a power-law continuum and an Fe template \citep{Bruhweiler2008} convolved with a Gaussian of variable width.}
\item{With the continuum and iron lines removed, the H$\beta$ line is fit with two Gaussians and the [O{\small III}]$\lambda 4959,5007$ doublet with a pair of Gaussians, with the [O{\small III}] doublet is constrained to have the 3:1 amplitude ratio physically required by the fine structure transitions involved and the two [O{\small III}] widths identical to the width of the H$\beta$ narrow component.  Typical uncertainties are 2-15\% in both H$\beta$ velocities, with more luminous quasars and broader components typically best determined.}
\item{The final result is checked for contamination by nearby He{\small I} lines.  If He{\small I} contamination is detected, we construct a new line fit with He{\small I} removed.  This check is not part of the \citet{Shen2008} fitting prescription.}
\end{enumerate}
The \citet{Shen2011} catalog includes best fits for objects with very poor signal to noise and low-amplitude lines.  We discard objects with very low H$\beta$ equivalent width and where a fit containing no H$\beta$ line had a better $\chi^2$/DOF than a fit containing an H$\beta$ line.  Although the Fe{\small II} width is often similar to the H$\beta$ broad component, they are not usually identical \citep{Hu2008b}.  As a result, this is not required in our fitting and for many AHQs, the two have very different velocities.  Because the Fe{\small II} amplitude is typically far smaller than the stronger of narrow H$\beta$ and [O{\small III}], alternative templates yield similar results.  A sample end result (the first well-measured example sorted by RA) is shown in Fig. \ref{fig:samples}.

In \citet{Shen2008}, the ``narrow'' component of H$\beta$ and [O{\small III}] are constrained to be narrower than 1200 km/s.  However, if H$\beta$ is indeed predominantly composed of two nearly-Gaussian components, one broader than 1200 km/s and another narrower than 1200 km/s and produced with the same peculiar velocity as [O{\small III}], this constraint is unnecessary, at least for quasar spectra with high signal-to-noise ratios. 

Surprisingly, though, this constraint turns out to have been quite important, because removing this constraint, while otherwise reproducing the technique used in the \citet{Shen2008} catalog, produces a substantial population of objects with ``narrow lines'' fit as broader than 1200 km/s.  In Fig. \ref{fig:samples}(bottom), the quasar SDSS J000011.96+000225.3 is shown to exhibit ``narrow'' lines with a FWHM of 1860 km/s.  Further, this is not an isolated example, but rather part of a large class of objects with a broadened H$\beta$ narrow component.  In the \citet{Shen2011} catalog, 30.4\% of objects with discernible H$\beta$ broad lines are measured to contain ``narrow'' lines broader than 1000 km/s.  It is convenient for the purposes of this paper to give this group of objects a label, although it is unclear what the proper nomenclature should be.  For reasons discussed in \S~\ref{sec:properties} and \S~\ref{subsec:nomenclature}, we will hereafter refer to this group of objects as ``correlated'' quasars, or AHQs.

\subsection{Separating [O{\small III}] and the H$\beta$ Core in AHQs}
\label{subsec:delink}

\begin{figure*}
  \vskip1.5cm
 \epsfxsize=3.25in\epsfbox{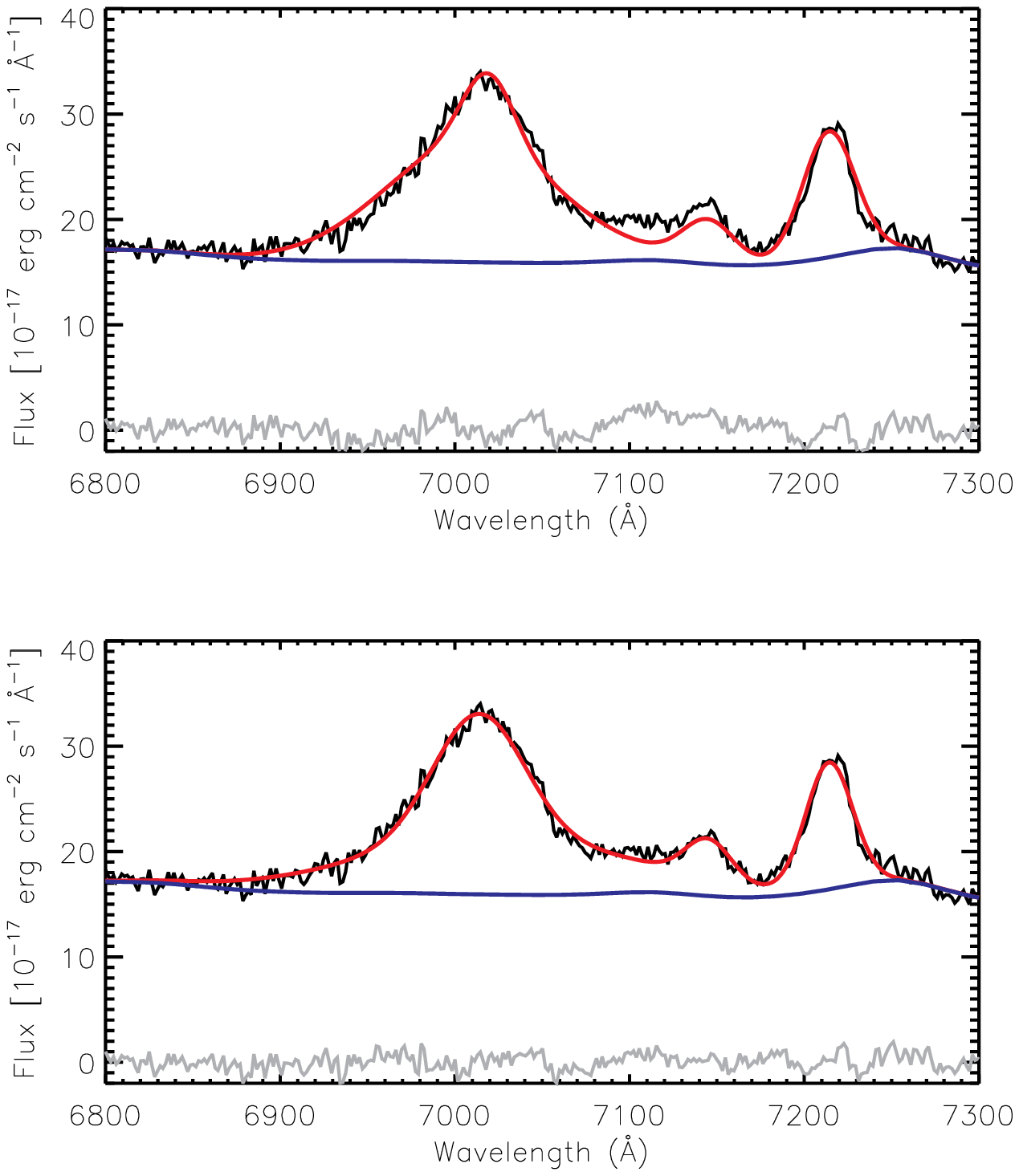}
  \caption{Best-fit H$\beta$ and [O{\small III}] lines for a sample AHQ.  At top, the best total fit (red) requiring that the [O{\small III}] line profile be identical to the H$\beta$ core, after subtracting the continuum and Fe{\small II} (blue).  At bottom, the best fit set of lines allowing the H$\beta$ core and [O{\small III}] to have separate profiles.  The de-linked fit is a better description of the observed spectrum, as evident both by eye and from the residuals (grey).}
  \label{fig:delink}
\end{figure*}

A closer examination of objects where these ``narrow'' lines are measured as being broader than 1000 km/s reveals a mismatch between the [O{\small III}] and H$\beta$ core profiles (Fig. \ref{fig:delink}).  
Fitting the [O{\small III}] line and the core of H$\beta$ independently for these objects produces a better fit.  Requiring [O{\small III}] and the H$\beta$ core to have identical profiles producs a 1940 km/s FWHM, but independent fits reveal a 2600 km/s FWHM for the H$\beta$ core and 1310 km/s for the pair of [O{\small III}] lines.  Although [O{\small III}] is narrower than the H$\beta$ core, after delinking the two lines, a population of objects with [O{\small III}] greater than 1200 km/s still exists.

It is expected that the H$\beta$ core should arise from the same narrow-line region as [O{\small III}]$\lambda 4959,5007$, and that they should therefore have the same profile.  There are a variety of different fitting methods for exploting that expected link, including fitting the [O{\small III}] profile independently and searching for an H$\beta$ component of the same width, looking for an H$\beta$ component with a profile fully identical to [O{\small III}], and the simultaneous fit for narrow H$\beta$ and [O{\small III}] used in \citet{Shen2008} and \citet{McLure2004}.  Surprisingly, we find that for AHQs, {\em none of these is a good description of the observed spectra} because the H$\beta$ core has a different, higher-FWHM profile than [O{\small III}].

It is likely that this broadened H$\beta$ core does not represent a broadened narrow-line component.  Typically the narrow-line region H$\beta$ component has a smaller amplitude than [O{\small III}], while AHQs are characterized by suppressed narrow line emission.  Rather, AHQs appear to contain two broad H$\beta$ components, both broader than narrow lines such as [O{\small III}], and no discernable narrow-line H$\beta$ component.  At the same time, [O{\small III}] can be broader than 1000 km/s, as in Fig. \ref{fig:delink}, and appears to be broader than 1000 km/s for approximately 10\% of the quasars in the SDSS catalog.

We note that although previous SDSS line catalogs \citep{Hao2005,Shen2008,Shen2011} have not reported a substantial sample of AHQs, objects with [O{\small III}] broader than 1000 km/s have been found in smaller surveys.  \citet{Marziani1996} found an [O{\small III}]$\lambda 5007$ FWHM over 1100 km/s for 6 of 52 low-redshift AGN.  In a sample of 32 high-luminosity quasars at $2.0 < z < 2.5$, higher redshifts than the AHQs reported in this work, \citet{McIntosh1999} found 12 with [O{\small III}]$\lambda 5007$ FWHM $\geq 1200$ km/s, and another six between 1000 and 1200 km/s, finding a strong correlation between [O{\small III}] FWHM and quasar luminosity.  \citet{Forster2001} point out that the average [O{\small III}] FWHM is 1150 km/s for radio-loud and 1160 km/s for radio-quiet quasars in the \citet{McIntosh1999} sample.  What we report is that these quasars with broadened [O{\small III}] are a part of a larger class of quasars with broader H$\beta$ cores than even these broadened [O{\small III}] lines.  

Although removing the link between the H$\beta$ core and [O{\small III}] produces a better description of the spectral line, for the remainder of this work (except for \S~\ref{subsec:delink2}) we will present fits linking the two lines.  We do so primarily for three reasons: 
\begin{enumerate}
\item{Standard line fitting techniques expect the H$\beta$ core and [O{\small III}] to have identical profiles.  Continuing to link the two lines demonstrates that our measurement of anomalous H$\beta$ cores does not arise from a fundamental change in fitting technique, but solely from removing the artificial constraint on H$\beta$ core velocity used in \citet{Shen2008}.}
\item{Because [O{\small III}] is narrower than the H$\beta$ core for AHQs, the combined fit underestimates the H$\beta$ core velocity.  Since this work focuses on reporting broad H$\beta$ cores, this underestimate yields a conservative AHQ fraction.}
\item{Existing catalogs, including SDSS value added catalogs \citep{Shen2011,Shen2008}, link the two lines.  Doing the same in our results below allows a direct comparison between our work and existing catalogs.}
\end{enumerate}
In short, future line catalogs for AHQs will likely avoid linking the H$\beta$ core and [O{\small III}] profiles.  However, because this work presents a surprising conclusion, it is presented using the most conservative technique available, underestimating both individual H$\beta$ core FWHMs and the overall AHQ fraction.  We also illustrate the effects of delinked line fitting on our main result in \S~\ref{subsec:delink2}.

\section{AHQ Properties}
\label{sec:properties}

The core of H$\beta$ is typically a narrow line and emitted from a region with low enough density to also emit forbidden lines such as [O{\small III}] and [Ne{\small V}], implying that these quasars with broader H$\beta$ cores (AHQs) might have atypical structure.  Thus, the investigation into AHQs begins by determining whether they are atypical in any other ways, perhaps accompanied by other distinguishing features that might help provide an explanation.

\begin{figure}
 \epsfxsize=3.25in\epsfbox{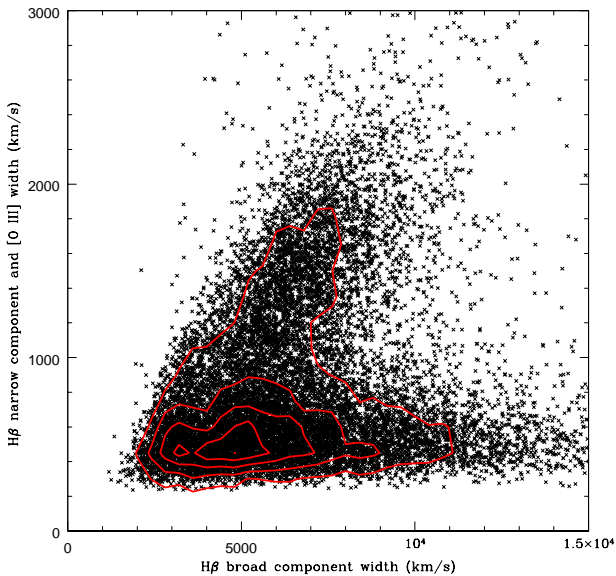}
\caption{Comparison of the narrow-line FWHM and H$\beta$ broad-component FWHM for quasars in the SDSS DR7 catalog.  Narrower velocities are accompanied by a wide range of H$\beta$ widths, but AHQ narrow-line velocities are well-correlated with broad H$\beta$, increasing with increasing broad H$\beta$ width.  Contours are drawn at equally-spaced number densities.}
\label{fig:linecomp}
\end{figure}

Indeed, a comparison between the broad H$\beta$ and narrower component FWHM shows two populations (Fig. \ref{fig:linecomp}), one with a narrow-line H$\beta$ core apparently uncorrelated with the H$\beta$ broad-line width and another with a broader H$\beta$ core that is well correlated with the broad H$\beta$ component.  It is unclear whether this correlation indicates two distinct, correlated physical regions or whether the correlation in Fig. \ref{fig:linecomp} is instead an non-physical artifact of the standard line fitting technique for H$\beta$ and [O{\small III}].  We consider this question and other fitting techniques further in \S~\ref{subsec:delink} and \S~\ref{subsec:delink2}.  What is clear, however, is that the standard fitting technique divides quasar emission profiles cleanly into two distinct groups, as in Fig. \ref{fig:linecomp}.  As we discuss in the remainder of this paper, the group of quasars on the ``correlated'' branch of Fig. \ref{fig:linecomp} have many other properties in common in addition to their H$\beta$ and [O{\small III}] profiles, and likely comprise a distinct set of quasars.  The question of whether this correlation actually indicates related physical emission regions or a different sort of gas dynamics likely requires additional observations and is beyond the scope of this paper.

The Type I quasar sample might thus be composed of two distinct classes of quasars: (1) quasars with standard narrow lines, of the type that has been previously studied and (2) quasars in which the physical process responsible for the anomalously broad H$\beta$ core also results in broader quasar ``broad lines'', which we describe as ``anomalous H$\beta$ quasars''.  Because low-density gas is required to allow the forbidden transitions producing the narrow [O{\small III}], the narrow-line region is often assumed to lie far from the central black hole.  A very strong outflow reaching the narrow line region could increase the velocity of [O{\small III}] (as is seen in some but not all AHQs) and the narrow component of H$\beta$, and would at the same time also increase the velocity of the broad H$\beta$ component coming from the broad-line region closer to the central black hole.

This second, diagonal branch of Fig, \ref{fig:linecomp} appears to include the ``intermediate line'' quasar sample previously reported by \citet{Hu2008} using a smaller sample of SDSS quasars.  \citet{Hu2008} found a similar fraction of AHQs to the $\sim 30$\% reported in our work.  By relating the AHQ branch to the standard quasars, we now find that these two classes may be more difficult to distinguish than previously thought.  As indicated in Fig. \ref{fig:linecomp}, a cut at $\sim 1200$ km/s is sufficient to remove standard quasars from an AHQ population, but objects with low FWHM of both H$\beta$ components, classed for our study as standard quasars, might instead lie on a continuation of the AHQ branch, or might be a combination of some objects with standard and some objects with AHQ physics in their broad-line regions.  If so, the true AHQ population would comprise more than 30\% of the SDSS catalog, and some AHQs would have been well-fit in the \citet{Shen2011} catalog.  This possibility is discussed further in \S~\ref{subsec:nomenclature}.

For further investigation, we divide the SDSS DR7 sample into bins by H$\beta$ core width, as in Table 1.
\begin{table*}
\begin{center}
\caption{Properties of quasars in the SDSS DR7 catalog binned by H$\beta$ core width, including the fraction showing broad absorption in Mg{\small II} (Shen et al. 2011) and the best-fit Fe{\small II} amplitude (arbitrary units).}
\begin{tabular}{|c|c|c|c|c|c|c|c|c|c|c|} 
\hline 
$\sigma$ (km/s) & FWHM & Color & N & Frac. & `Best' N & Frac. & BAL & Fe{\small II} (arb.) & $\log L_{bol}$ \\
\hline 
100--200 & 235--471 & Black & 3549 & 0.217 & 1162 & 0.217 & 0.0033 & 1.59 & 45.41 \\
200--300 & 471--706 & Black & 4941 & 0.303 & 1018 & 0.274 & 0.0050 & 2.14 & 45.51 \\
300--400 & 706--942 & Red & 2309 & 0.141 & 657 & 0.123 & 0.0098 & 3.25 & 45.61 \\
400--500 & 942--1177 & Red & 1366 & 0.084 & 420 & 0.078 & 0.0069 & 4.14 & 45.68 \\
500--600 & 1177--1413 & Yellow & 1086 & 0.067 & 394 & 0.074 & 0.0084 & 7.10 & 45.72 \\
600--700 & 1413--1648 & Green & 1018 & 0.062 & 432 & 0.081 & 0.0065 & 9.37 & 45.82 \\
700--800 & 1648--1884 & Cyan & 743 & 0.045 & 321 & 0.060 & 0.0116 & 10.48 & 45.90 \\
800--900 & 1884--2119 & Blue & 424 & 0.026 & 207 & 0.039 & 0.0249 & 11.96 & 46.00 \\
900--1000 & 2119--2355 & Magenta & 227 & 0.014 & 110 & 0.021 & 0.0284 & 13.75 & 46.01 \\
\hline  
\end{tabular}
\end{center}
\label{table:bins}
\end{table*}
Quasars with standard narrow lines are most common, with the population peaking near the expected 500 km/s line FWHM.  However, 30.4\% of quasars have H$\beta$ core velocities greater than 1000 km/s, and in nearly one quarter of quasars these velocities are above the \citet{Shen2008} limit of 1200 km/s.  Although Fig. \ref{fig:linecomp} indicates that there are two populations of quasars, the quasar fraction decreases monotonically at velocities above the peak.  This may indicate that quasars are not necessarily either AHQs or ``standard'' quasars, but rather can also exist in an intermediate state.  Anomalous narrow lines are accompanied, on average, by an increased luminosity.  However, AHQs are also increasingly likely to contain broad absorption lines as measured from Mg{\small II} in the \citet{Shen2011} catalog.

\begin{figure}
 \epsfxsize=3.25in\epsfbox{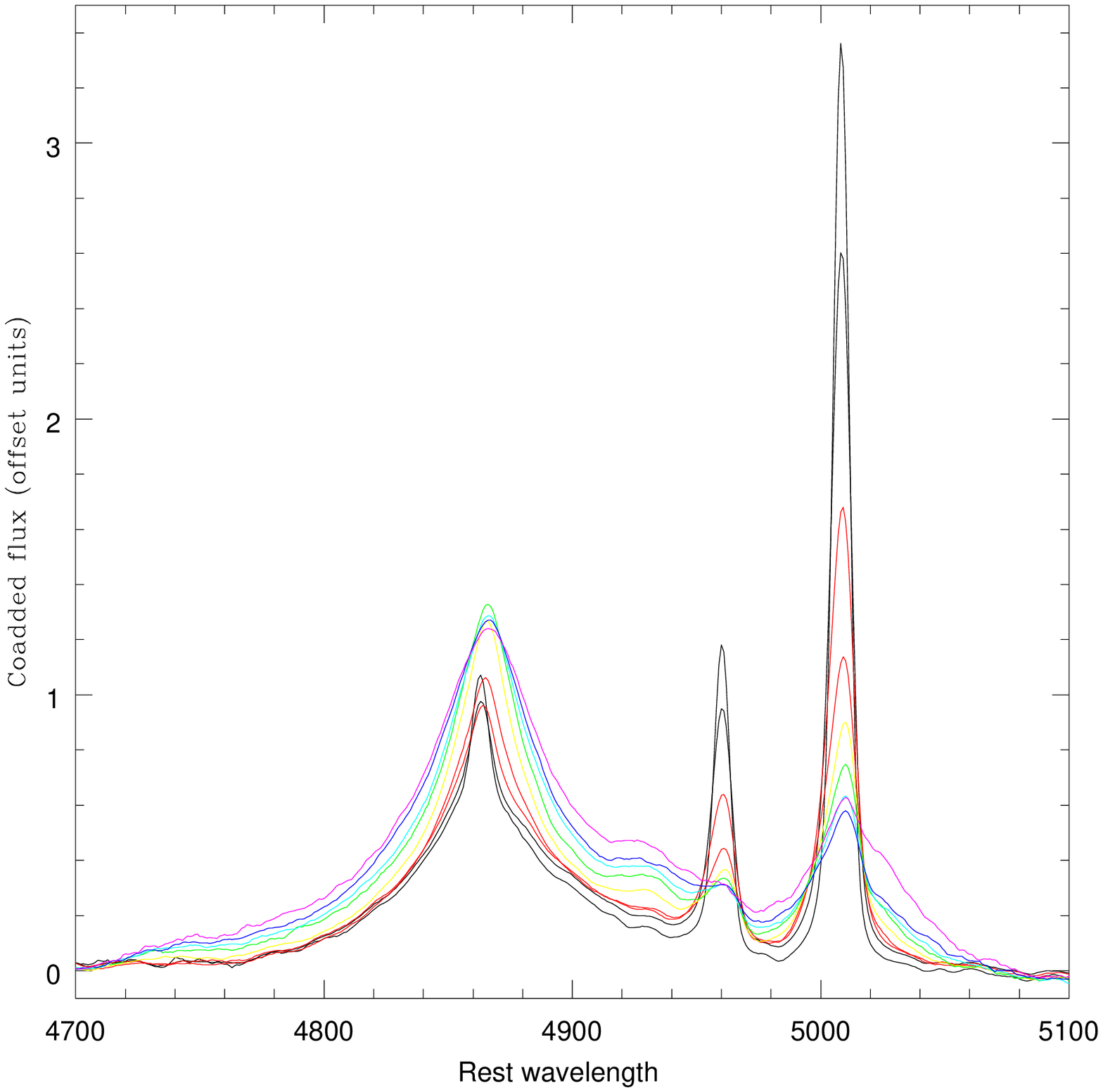}
\caption{Comparison of co-added quasar spectra around the H$\beta$ and [O{\small III}] lines binned and colored as in Table 1.  The continuum and Fe{\small II} lines have been subtracted.}
\label{fig:hboiiistack}
\end{figure}

For a variety of reasons, our best fit may not be a good match for even a well-measured SDSS spectrum, and this is one of the limitations of any automated line-fitting prescription with a small number of free parameters.  For well-measured spectra where narrow H$\beta$ and [O{\small III}] have similar, Gaussian profiles, our fit is generally good.  However, many quasars best-fit with standard ``narrow'' lines have non-Gaussian profiles, and many anomalous H$\beta$ quasars show a mismatch between the H$\beta$ and [O{\small III}] profiles (\S~\ref{subsec:delink}).  In such cases, our fitting routine linking the two will typically be dominated by the higher-amplitude of narrow H$\beta$ and [O{\small III}], which typically is [O{\small III}] for quasars with a standard set of narrow line and H$\beta$ for AHQs (Fig. \ref{fig:hboiiistack}, further described below).  

Some AHQs showing a mismatch between the H$\beta$ core and [O{\small III}] may be objects with a broad blue [O{\small III}] wing component tied to the H$\beta$ core and in addition to the O{\small III} narrow component, similar to those discussed in \citet{Zhang2011}.  Quasars with $\sim 2000$ km/s H$\beta$ cores and standard [O{\small III}] lines are certainly anomalous, but might have a different cause than AHQs in which [O{\small III}] also has a discernable broad component.

We produce a `best' sample of quasars for which (1) the overall fit $\chi^2/\textrm{DOF}$ is good and (2) the $\chi^2/\textrm{DOF}$ for the $\sim 50-150$ pixels within 2$\sigma$ of the H$\beta$ core and [O{\small III}] is within 0.5 of the total $\chi^2/\textrm{DOF}$.  As in Table 1, this sample has a slightly larger AHQ fraction than the overall catalog.  A very conservative lower bound on the true AHQ fraction would be 11\%, the fraction of quasars out of the entire sample that are AHQs in the `best' sample.  A much more likely explanation is that $\sim 11$\% of quasars are AHQs in which the broadened H$\beta$ core profile is accompanied by a somewhat broadened [O{\small III}] profile, while an additional $\sim 20$\% of quasars are AHQs with H$\beta$ broadened but [O{\small III}] suppressed and unbroadened.

Spectra within each bin from Table 1 are co-added to examine the average dependence of spectral lines on H$\beta$ core FWHM.  Each individual spectrum is first smoothed to the resolution of the SDSS spectrograph, then normalized to the monochromatic 5100 \AA~flux and averaged.  Subtracting the best-fit continuum and Fe{\small II} template results in the spectral lines shown in Fig. \ref{fig:hboiiistack}.
As indicated by Fig. \ref{fig:linecomp}, for typical H$\beta$ core profiles (black, red) in Fig. \ref{fig:hboiiistack} an increase in FWHM is not associated with strong changes in the broad component of H$\beta$.  However, broader H$\beta$ core profiles, as in AHQs (remaining colors), are associated with an increase in width of H$\beta$ as well as increasing H$\beta$ line flux.  When the [O{\small III}] width increases, it is associated with a declining amplitude, and in total a nearly-constant equivalent width.

\begin{figure*}
 \epsfxsize=7in\epsfbox{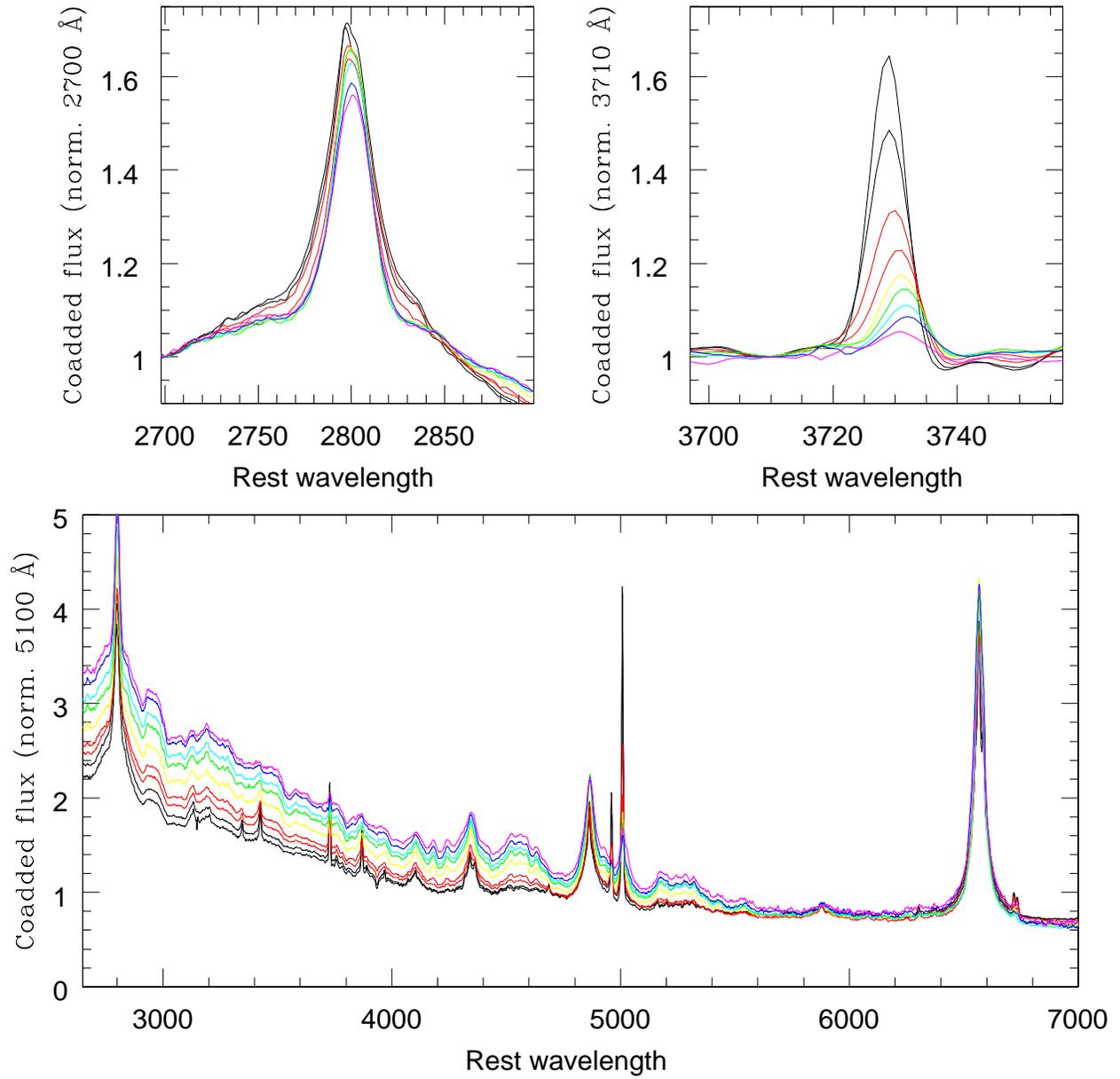}
\caption{Comparison of co-added quasar spectra at different narrow-line widths binned and colored as in Table 1.  For shorter wavelengths, only spectra at sufficiently high redshift were co-added.  At top left, the Mg{\small II} region.  At top right, the [O{\small II}] region.}
\label{fig:combo}
\end{figure*}

Several other spectral features are also responsive to increased H$\beta$ core width (Fig. \ref{fig:combo}).
The continuum tilt increases for AHQs, making them bluer than other quasars.  This bluer tilt occurs despite a higher BAL fraction among AHQs (Table 1), which indicates AHQs might be on average dustier than other quasars.  Perhaps this continuum slope is explained by AHQs having a higher temperature, also consistent their higher average luminosity.  AHQs also show stronger Fe{\small II} lines (see also \citet{Boroson1992}, Table 1).  An effort to find AHQ indicators available at other redshifts where [O{\small III}] is unavailable finds that [Ne{\small V}] broadens along with [O{\small III}], while [S{\small II}] does not \citep{Steinhardt2011b}.

The stronger and broader H$\beta$ line might be evidence that the nature of the quasar broad-line region is changing, whether due to a strong outflow or other accretion physics.  However, the Mg{\small II} line (Fig. \ref{fig:combo}) does not increase in width or in equivalent width in AHQs.  Since Mg{\small II} has an ionization potential placing it at a larger radius from the central black hole \citep{Petersonbook}, one possible explanation might be an outflow propagating only partway through the broad-line region. 

However, the [O{\small II}] $\lambda 3727$ \AA~line appears sensitive to these changes (Fig. \ref{fig:combo}).  [O{\small II}] originates in both narrow-line region gas around the quasar and star formation in the host galaxy \citep{Ho2005,Kim2006}.  If broadened [O{\small III}] seen in some AHQs is evidence of an outflow broadening the entire narrow-line region, then any [O{\small II}] coming from near the quasar would also be broadened.  The [O{\small II}] amplitude is declining in Fig. \ref{fig:combo}, but the line width is not increasing.  

The decline in [O{\small II}] amplitude indicates that the combination of [O{\small II}] from all sources, including star formation, diminishes with increasing H$\beta$ core width.  One intriguing possibility is that star formation in AHQ hosts is quenched by outflows from the central black hole.  It is also possible that all quasars inhibit star formation, and that the reduction in [O{\small II] flux between standard Type I and AHQ quasars is due to a reduction in the photoionized [O{\small II}] in the narrow-line region.

The [O{\small II}] centroid appears to shift between different bins in Fig. \ref{fig:combo}.  Although this could be interpreted as a skew in [O{\small II}], the redshift is determined primarily by the highest-equivalent width lines.  If H$\beta$ is skewed in AHQs (as indicated in Fig. \ref{fig:hboiiistack}), the result may be a systematically incorrect redshift, resulting in locating low-amplitude lines such as [O{\small II}] at the wrong rest wavelength.  If the narrow component is blueshifted while the broad component lies at the velocity of the host galaxy, it might induce this sort of skew.  This effect may make high-precision redshift determination difficult for AHQs, as their broad emission lines are skewed and their galactic lines are difficult to measure due to their small amplitude. 

If this redshift error leading to an [O{\small II}] offset is systematic and uniform for fixed narrow-line width, the average decrease in [O{\small II}] amplitude will be well-indicated by the co-added spectrum.  However, if the offset is non-uniform, high-amplitude lines with different centroids may be co-added to produce a low-amplitude, broader line.  Therefore, we fit the [O{\small II}] line directly and investigate its properties.  As with our H$\beta$ and [O{\small III}] fits, we first fit a combination of continuum and broadened Fe{\small II} template, then find the best-fit Gaussian for the leftover line profile.  We find that the [O{\small II}] equivalent width declines by a factor of 2.25 for AHQs compared to standard quasars.  The best-fit [O{\small II}] width remains typical of a narrow galactic emission line, and the decreased line flux comes from to a decrease in [O{\small II}] amplitude.  For low-amplitude [O{\small II}], noise is often incorporated into the best-fit line profile, so the true decline may be larger. 

\section{Dependence of AHQs on Fitting Techniques}
\label{sec:fitting}

Because of the long history of studying quasar spectra, a report of a large, new spectroscopic class of quasars discovered via spectral line fitting should be met with appropriate skepticism, and it is important to consider whether AHQs might just be an artifact of fitting techniques.  Ultimately, the H$\beta$-[O{\small III}] complex arises from complicated physics but is fit with a small number of parameters in an automated way, without the underlying model being adjusted individually for tens of thousands of quasars in the SDSS DR7 and future catalogs.  As a result, most fits should contain errors.  By choosing a technique previously used by \citet{McLure2004} and \citet{Shen2008}, we hope to have avoided introducing any novel line-fitting errors.  However, we should highlight two specific dangers:

\subsection{Continuum Fitting}

Errors in continuum fitting could result in incorrectly determining the properties of broad spectral lines, either missing the broad component of H$\beta$ if too much flux is considered part of the continuum or introducing a false broad component by under-fitting the continuum.  Although for $\sim 500-1000~\AA$ windows the continuum is well-fit by a power law, the slope of that power law changes between the Mg{\small II} and H$\beta$ region in composite spectra \citep{VandenBerk2001}, so it must be locally determined.  However, most of the pixels near H$\beta$ are contaminated by either Fe{\small II} or strong spectral emission lines.  Our fitting technique solves this problem by fitting simultaneously for Fe{\small II} emission and the continuum.  An alternative technique would involve choosing a window free of Fe{\small II} contamination, such as 5080--5100~\AA \citep{Forster2001}, and using these windows to fit the continuum alone.  However, in an AHQ with broadened [O{\small III}], 5080~\AA may not be free of [O{\small III}] contamination, so this window could also provide an erroneous continuum slope.  

By simultaneously fitting Fe{\small II} and continuum, we fit hundreds of SDSS pixels with four parameters, resulting in a well-determined fit.  However, although the continuum fits exhibited in Figs. \ref{fig:samples} and \ref{fig:delink} appear reasonable, it is difficult to prove that any individual spectrum has been well-fit.  Fortunately, it is the broad, rather than the narrow, component of H$\beta$ that will be strongly affected by errors in continuum fitting.  Our report of a large population of AHQs relies on fitting the H$\beta$ narrow component, and thus should not be susceptible to errors in continuum fitting.

\subsection{Iron Contamination}

Although Fe{\small II} is a larger problem when fitting Mg{\small II} rather than H$\beta$, there are also Fe{\small II} lines extending near [O{\small III}].  Further, the strength of iron lines increases strongly in AHQs (Table 1). 
As with continuum fitting, Fe{\small II} contamination should not be a problem when determining whether an object is some type of AHQ, as the narrow H$\beta$ component is stronger than Fe{\small II} and well-measured.  However, the combination of strongly suppressed [O{\small III}] and augmented Fe{\small II} emission is more likely to result in errors fitting [O{\small III}]$\lambda 5007$.  

The presence of stronger iron lines does make Fe{\small II} easier to detect and fit.  Further, although there are several different Fe{\small II} templates available, they are typically derived from the spectra of objects with narrower Fe{\small II} lines, such as 1 Zw 1 \citep{Bruhweiler2008}.  Because these templates are then convolved with a broad Gaussian to simulate Fe{\small II} with a FWHM of thousands of km/s, differences between templates are minimized, so alternative templates produce a similar AHQ sample.

The reduction from the full AHQ sample to the `best' sample presented in \S~\ref{sec:properties} uncovered a large population of AHQs in which [O{\small III}] was suppressed but not broadened.  Perhaps what we have fit as suppressed and broadened [O{\small III}] in these other cases is instead dominated by strong Fe{\small II} emission at broad-line region velocities.  Because several AHQs show not just a broadened [O{\small III}]$\lambda 5007$ line but also equally-broad [O{\small III}]$\lambda 4959$, for some objects [O{\small III}] is unambiguously broader than 1000 km/s (Fig. \ref{fig:delink}).  In other objects, a combination of He{\small I} contamination and highly-suppressed [O{\small III}]$\lambda 4959$ might make an alternative explanation involving Fe{\small II} more plausible.

The Fe{\small II} FWHM is often linked to the broad H$\beta$ component FWHM.  However, in AHQs with broadened [O{\small III}], the [O{\small III}] FWHM is instead linked to the H$\beta$ narrow component, an apparent mismatch.  Because AHQ broad-line regions appear to behave anomalously in several ways compared to BLRs in standard quasars, perhaps iron emission is also different.  Iron contamination is therefore a plausible concern when fitting [O{\small III}] for some AHQs, but cannot be responsible for miscategoring objects as AHQs based upon the core H$\beta$.  
 
\subsection{De-linking [O{\small III}] and H$\beta$}
\label{subsec:delink2}

As described in \S~\ref{subsec:delink}, one way in which the standard fitting technique used in this paper seems clearly inadequate for AHQs lies in linking the core H$\beta$ profile with the profile of nearby [O{\small III}] lines.  For the reasons given earlier, we have presented our main results using a more standard fitting technique.  For quasars with a standard narrow-line region, the [O{\small III}] and H$\beta$ core profiles are indeed similar, so that de-linking them has a negligible effect.  However, for AHQs de-linking the two profiles often results in the detection of broader H$\beta$ core than the [O{\small III}] width.  

Using a linked fit, AHQs appeared as a diagonal branch comprising at least 30\% of quasars (Fig. \ref{fig:linecomp}).  By de-linking the H$\beta$ narrow component from [O{\small III}] (Figs. \ref{fig:delinkhb}-\ref{fig:delinkoiii}), we find that for some quasars, both the H$\beta$ core and [O{\small III}] are broadened, but there are also many objects for which the H$\beta$ core is broadened but [O{\small III}] is at a typical narrow-line velocity of $< 1000$ km/s.
\begin{figure}
 \epsfxsize=3.25in\epsfbox{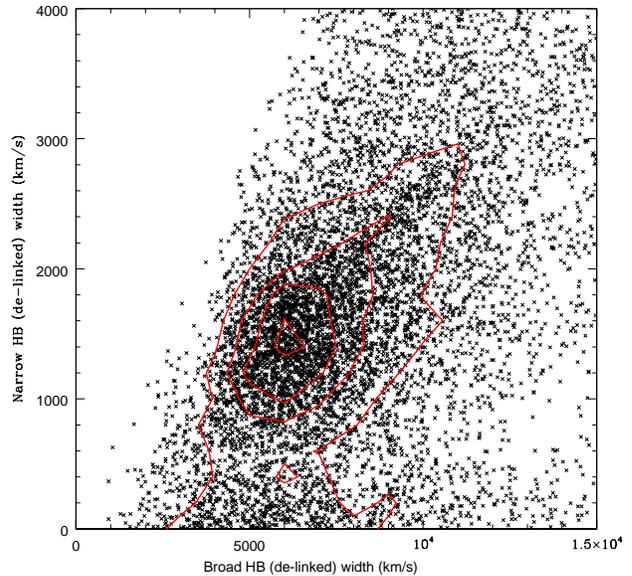}
\caption{{Comparison of the H$\beta$ narrow-component FWHM and H$\beta$ broad-component FWHM for quasars in the SDSS DR7 catalog, using a fitting technique in which the H$\beta$ profile is not linked to other narrow lines such as [O{\small III}].  Narrower velocities are accompanied by a wide range of H$\beta$ widths, but AHQ narrow-line velocities are well-correlated with broad H$\beta$, increasing with increasing broad H$\beta$ width.  Contours are drawn at equally-spaced number densities.  Approximately 0.1\% of two-component fits showed nearly identical widths and were rejected due to degeneracy.}}
\label{fig:delinkhb}
\end{figure}
\begin{figure}
 \epsfxsize=3.25in\epsfbox{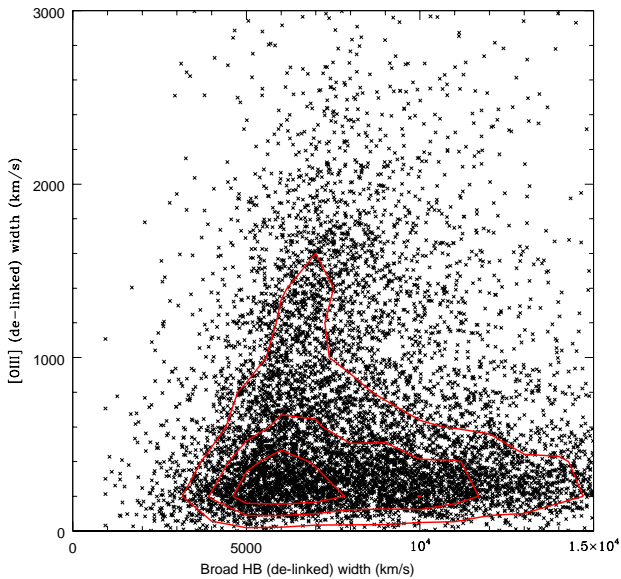}
\caption{{Comparison of the [O{\small III}]$\lambda$4959,5007 FWHM and H$\beta$ broad-component FWHM for quasars in the SDSS DR7 catalog, using a fitting technique in which the H$\beta$ profile is not linked to other narrow lines such as [O{\small III}].  Narrower velocities are accompanied by a wide range of H$\beta$ widths, but AHQ narrow-line velocities are well-correlated with broad H$\beta$, increasing with increasing broad H$\beta$ width.  Contours are drawn at equally-spaced number densities.}}
\label{fig:delinkoiii}
\end{figure}

That there is a population of AHQs when fitting the [O{\small III}] line independently (Fig. \ref{fig:delinkoiii}), similar to Fig. \ref{fig:linecomp} demonstrates that the broadened [O{\small III}] in AHQs is not simply an artifact of tying the [O{\small III}] profile to H$\beta$.  Further, in quasars with broadened [O{\small III}], its velocity is correlated with broad H$\beta$.  Thus, the apparent connection between the dynamics of gas emitting [O{\small III}] and broad-line region gas strongly influenced by the central black hole must also be physical rather than an artifact of our line-fitting techniques.  We note that for all other quasars, the de-linked fits produce the standard result that the dynamics of the narrow-line region are (nearly-)independent of those of the broad-line region.

For the H$\beta$ line, de-linking the fits also provides a similar picture to Fig. \ref{fig:linecomp}, but now places far more quasars on the AHQ branch in Fig. \ref{fig:delinkhb}.  Since Fe{\small II} contamination is a concern when fitting the wings of the [O{\small III}] profile but not for H$\beta$, we conclude that Fe{\small II} cannot provide an explanation for AHQs.  We also note that Fig. \ref{fig:delinkhb} is similar for different flux ratios between the two H$\beta$ components, including where the narrower component has similar flux to the broader component and where there is a large disparity.  

Because  [O{\small III}] is typically stronger than the H$\beta$ core, the FWHM for the linked fit is usually dominated by [O{\small III}], and thus quasars in which the H$\beta$ core is broad but [O{\small III}] is narrow appear on the standard quasar branch in Fig. \ref{fig:linecomp}.  Our estimate using linked fitting that approximately 30\% of quasars are AHQs therefore is an underestimate for H$\beta$, and instead is closer to the fraction of quasars with both broadened H$\beta$ cores and [O{\small III}].  

Finally, we note that even when fitting [O{\small III}] independently of the H$\beta$ profile, there appear to be two distinct branches.  It appears that when H$\beta$ has a broadened core, [O{\small III}] emission is possible at a typical narrow-line velocity, possible at the H$\beta$ core velocity, but impossible at a velocity in between those two.  When H$\beta$ is well-described by two tightly-correlated components, it is unclear whether this indicates two distinct regions with correlated dynamics or one region with a complex profile that can be fit with two correlated Gaussians.  This evidence from [O{\small III}] seems to indicate that when H$\beta$ has a broadened core, the core emission may physically arise from a distinct region, one which sometimes lies at a low enough density to allow [O{\small III}] emission.

\subsection{Modeling H$\beta$ With Two Gaussians}
In this work, we choose to model the H$\beta$ line as a pair of Gaussians, and each of the two [O{\small III}] lines as one Gaussian.  This is in keeping with standard practice for large catalogs (cf. \citet{Shen2011,Shen2008,McLure2004,Rafiee2011}, etc.).   Even some past work using more complicated fitting prescriptions reported their results using two Gaussians \citep{Hu2008}.  As such, if the correct interpretation of this result turns out to be merely that the two-Gaussian fit is faulty for $\sim 30$\% of quasars, it would still be important to understand and correct for this problem, and that correction would change the results of existing statistical studies of the SDSS quasar population using H$\beta$ line fitting, including virial mass catalogs. 

However, the question remains as to whether two Gaussian components is a good representation of the H$\beta$ line, particularly for AHQs.  When only one component of an AHQ is allowed to be broader than 1200 km/s, effectively the entire line shape is modeled as one Gaussian, where the fitting in this work uses two.  Fitting two broad Gaussians to a line shape closer to one Gaussian, but not exactly Gaussian, might lead to fit composed of the dominant Gaussian and a second, correlated Gaussian that would be artificial and representative of line non-Gaussianity.  So, is one Gaussian actually a better fit for AHQs?  For that matter, is a third Gaussian an even better fit, as reported by \citet{Hu2008}? 

The simplest measure of whether two Gaussians provide a better fit is to examine the $\chi^2$/DOF for each fit.  We consider the part of the spectrum within $2\sigma$ of the peak, as defined by the broadest component.  For all quasars, the $\chi^2$/DOF averages 1.31 for two-Gaussian fits and 1.70 for one-Gaussian fits.  For AHQs, two-Gaussian fits average 1.30 while one-Gaussian fits average 2.02.  Adding a third Gaussian, however, provides no such substantial help; a three Gaussian fit reduces the $\chi^2$/DOF from 1.31 to 1.29 for all quasars and, like the two-Gaussian fit, averages 1.30 for AHQs.  Thus, it is clear that the second-strongest component of the H$\beta$ lineshape is strong, while the third-largest is much weaker.  A two-Gaussian fit therefore seems reasonable, although the correct line shape may not contain Gaussians at all; a correct fit should produce a $\chi^2$/DOF of 1.  However, errors in SDSS are likely correlated between neighboring pixels \citep{McDonald2006}, so the reported uncertainties used in producing these value for $\chi^2$ may not be trustworthy.  This $\chi^2$ analysis is a good sanity check on our fitting technique, but is certainly not proof that the lines are being fit well.

Ultimately, fitting lines is an art, there are many complex prescriptions that can be used to produce higher-quality fits for individual spectra, and it is unrealistic to tailor individual fits for 100,000 spectra in SDSS, or even more in future surveys.  We believe the best we can do as scientists is to use public data and encourage other experts to use alternative prescriptions for H$\beta$.  It appears very difficult to fit AHQs well with one broad component and one component under 1200 km/s in FWHM.

In general, there are many errors arising from fitting the spectrum of a complex object with a simplistic model containing little physics and a small number of parameters.  These errors will likely result in some incorrect fits, and it would be difficult to believe reported H$\beta$ core widths to within 1\%.  However, many AHQs have H$\beta$ cores with FWHM measured at over 2000 km/s, and even a rudimentary fit ``by eye'' to the H$\beta$ lines displayed in Figs. \ref{fig:samples} and \ref{fig:delink} cannot accommodate more typical H$\beta$ core FWHM of under 1000 km/s.  These details of line-fitting are important when, for example, using precision line measurements to produce a quasar virial mass.  However, they cannot be responsible for overestimating the H$\beta$ core width by such a large factor in 30\% of the SDSS quasar catalog.

\section{Discussion}
\label{sec:discussion}

We report a new class of Type I quasars with broadened H$\beta$ cores.  These quasars (AHQs) include over 1/4 of all $\sim 16000$ SDSS quasars for which H$\beta$ is well-measured ($z < 0.8$).  However, previous line-fitting catalogs of SDSS quasars have not reported finding AHQs for two reasons: (1) because selection and fitting routines currently assume that the H$\beta$ core has the same profile as [O{\small III}], and that both are narrow lines, which is not true for AHQs, and (2) because H$\beta$ lines that cannot be fit well be one broad and on narrow component have been modeled with additional Gaussians and more complex profiles \citep{Shen2011}, rather than simpler, two-component model used in this work.

The existence of these AHQs now necessitates a re-examination of other selection criteria in SDSS and similar surveys.  In particular, the presence of a strong, narrow [O{\small III}] line is often a strong component of Type II quasar and Seyfert galaxy selection, since broader lines may not be present.  Any type II AHQs or AHQ Seyferts would likely be selected against during the production of such catalogs.

Several previous examples have been reported of other AGNs with anomalous narrow lines.  These include Seyferts with broadened or skewed [O{\small III}] (cf. \citet{Xu2011}), as well as quasars with multiple broad H$\beta$ components \citep{Hu2008}, with the broader broad component redshifted with regard to the narrower one \citep{Sulentic2002,Collin2006}.  

AHQs, however, not only show two well-centered broad components but represent approximately one quarter of the SDSS quasar catalog at $z < 0.8$, with thousands of examples.  As such their origin is a compelling puzzle.  The spectra of these quasars provide enough evidence to analyze several possible explanations:

\noindent{\bf Could AHQs be created by an outflow into the narrow-line region?}  Given that outflows are a common property of most quasars \citep{Elvis2000}, it would not be surprising to find that a large population exists with discernable evidence for possible non-virial gas motion in the broad-line (and narrow-line) region.  Although there is a slight skew to [O{\small III}] and the narrow component of H$\beta$ in co-added AHQ spectra (Fig. \ref{fig:hboiiistack}), the lines are mostly symmetric.  Thus, any outflow must be similarly symmetrical, meaning that light from both sides of the outflow is visible.  Otherwise, these lines would be more strongly skewed, just as strong asymmetric outflows can skew C{\small IV} emission \citep{Shen2008}.  

Moreover, Mg{\small II}, which emanates from the outer portion of the broad-line region, does not increase in width in AHQs (Fig. \ref{fig:combo}).  The narrow-line region, containing [O{\small III}] and presumably the H$\beta$ core, lies further from the central black hole than the broad-line region.  Thus, since any outflow does not extend past the edge of the broad-line region, AHQs are not created by an outflow into the narrow-line region.  The broad component of H$\beta$ also comes from the broad-line region, but its ionization potential places it closer to the central black hole than Mg{\small II}.  Therefore, an outflow that does not reach Mg{\small II}-emitting gas could still be responsible for higher velocities in gas producing the broad component of H$\beta$.  An alternative option might be an outflow, larger than the accretion disk, that is able to emit H$\beta$ (and, presumably, C{\small IV}), but unable to emit Mg{\small II}.

\noindent{\bf Could AHQs be created by H$\beta$ core emission from virialized gas in the broad-line region?}  Some AHQs contain not just a broadened H$\beta$ core, but also show [O{\small III}] broader than  1000 km/s (Fig. \ref{fig:delink}).  As a forbidden line, [O{\small III}] requires a low density.  In a typical Type I quasar, the broad-line region is too dense to allow [O{\small III}] emission, which is why [O{\small III}] does not have a prominent broad-line component.  In AHQs, the line flux from H$\beta$ and Mg{\small II}, both emitted in the broad-line region is comparable to or larger than other Type I quasars.  So, the broad-line region density is unlikely to be substantially lower.  Further, if [O{\small III}] and the narrow component of H$\beta$ are in the broad-line region, where is the broad component of H$\beta$ located?  Nevertheless, many AHQs show a broadened H$\beta$ core and suppressed [O{\small III}] emission, a combination more suitable for broad-line emission.

One way to test this picture is to consider the evidence from virial mass estimates.  As described in \citep{Steinhardt2011b}, the H$\beta$ broad component yields a systematically larger mass estimate than Mg{\small II} for AHQs.  Unlike H$\beta$, Mg{\small II} is not strongly correlated with changes in [O{\small III}] for AHQs, and therefore Mg{\small II} is likely the better mass indicator.  For agreement, the H$\beta$ broad component must be placed at less than half of the radius inferred from the continuum luminosity - broad line region radius relation calibrated by reverberation mapping.  If the H$\beta$ narrow component and [O{\small III}] were instead placed at that radius, it would result in virial masses 1-2 dex below those produced using Mg{\small II}.  Thus, H$\beta$ core emission from virialized broad-line region gas does not seem to fit the available evidence, although some of that evidence may not be reliable.

\noindent{\bf Could AHQs be typical Type I quasars viewed only from certain angles where an outflow is seen most clearly?}  [O{\small II}] emission comes in part from star formation and in part from larger radii than [O{\small III}], the latter due to its lower ionization state.  Thus, [O{\small II}] emission sharply declining (Fig. \ref{fig:combo}) with broadened H$\beta$ cores seems to indicate that AHQs lie in host galaxies, or at least $\sim$ kpc scale central regions, with different properties than other Type I quasars.  This would not happen merely due to geometry.  Indeed, it is a puzzle that the [O{\small II}] line indicates a connection between the host galaxy and the quasar, while the unchanging Mg{\small II} indicates that the connection is not directly due to a direct outflow.

\noindent{\bf Could AHQs be a pre-turnoff phase of the quasar duty cycle?}  The decreased [O{\small II}] equivalent width, coming from a combination of star formation and narrow-line region gas, might indicate a reduced supply of gas and dust to the central black hole, perhaps further reduced by a strong outflow.  This, in turn, could cause the central quasar to go quiescent, with AHQs an intermediate phase as the quasar turns off.  However, the average luminosity of AHQs is actually higher than other quasars (Table 1).  An upcoming luminosity decrease would not necessarily render the quasar quiescent, and the response time to a reduced inflow $\sim$ kpc away is long.  Nearby, AHQs might actually have more dust available than average Type I quasars, as they are more likely to contain broad absorption lines (Table 1).  This picture also does not yet explain the broadened H$\beta$ and [O{\small III}] lines.

So, although several scenarios present themselves, none yet seems able to fit all of the available evidence.  

\subsection{AHQs and Feedback}

There is known to be a link between the properties of lower-luminosity active galactic nuclei and their hosts \citep{Kauffmann2003,Silverman2009,Schawinski2010,Cardamone2010}.  Although co-evolution of the quasar and its host galaxy has long been expected and even assumed, direct evidence for this idea has not yet been forthcoming, primarily because the quasar dominates its host and because the peak quasar number density lies at $z \sim 2$, where resolving the host galaxy as separate from the quasar cannot be done with current ground-based instruments and where the quasar itself, the accretion disk and the broad-line region cannot be resolved with any existing instruments.  

One particularly striking feature of AHQ quasars is the strong link in co-added AHQ spectra (Fig. \ref{fig:combo}) between [O{\small III}], produced either in nearby heated interstellar medium or closer to the central black hole, and [O{\small II}], produced in photoionization $\sim$ kpc from the black hole and in star formation in the host galaxy.  The correlation between AHQs and their hosts appears to be evidence of some sort of feedback or linked evolution.  Further work, including a physical model for that link, is needed to understand whether we are seeing feedback, co-evolution, perhaps some other process that, as a byproduct, quenches star formation, broadens H$\beta$ core, and sometimes even broadens the [O{\small III}] line.

\subsection{Nomenclature}
\label{subsec:nomenclature}

One of the difficulties in reporting on quasars with anomalous H$\beta$ cores has come in choosing the terminology, particularly for objects in which [O{\small III}] is also broader than 1000 km/s.  In many ways, a better description might invoke an unsatisfying combination such as ``anomalous narrow-line'', ``broad narrow-line'', or even ``broad narrow H$\beta$'', in describing these quasars.  Alternatively, perhaps they might be named in analogy with Seyfert nomenclature (e.g., NLSy1).  It was even pointed out to the authors that with AHQs comprising as much as 30\% of the quasar catalog, their H$\beta$ profiles aren't really anomalous.  The root problem is that a set of lines conventionally referred to as ``narrow'' are often found at velocities typically termed ``broad'', and it may be difficult to satisfactorily describe these objects without altering common terminology.  We have chosen to describe these objects as ``anomalous H$\beta$'' in an effort to avoid this debate, and it is our hope that future work will produce a physical understanding of AHQs and allow a name representative of their origin rather than their spectra.

It should also be noted that the selection of AHQs in this study has been based entirely upon spectroscopic criteria, i.e., their H$\beta$ narrow component is anomalously broad.  A more natural categorization might be based upon the physics of their broad-line regions.  For example, quasars might be classed as either virial (VQ) or non-virial (NVQ) based upon their broad-line regions.  As discussed in followup work, many AHQs appear to have non-virial broad-line regions \citep{Steinhardt2011b}.  One might guess that quasars on the diagonal branch of Fig. \ref{fig:linecomp} might be NVQs, while quasars on the horizontal branch are VQs, with the quasars in the lower-left intersection of the two branches perhaps consisting of a mixture of both VQs and NVQs.  In this interpretation, what we have currently labeled our AHQ sample would consist most of NVQs with a few VQs mixed in, while our ``standard'' quasar sample would consist mostly of VQs but with a subsantial number of NVQs as well at lower-left.  

This work has defined AHQs as a population for which standard assumptions about the broad- and narrow-line regions appear to be false.  This is a useful definition because line-fitting catalogs are often predicated upon these assumptions, and for AHQs an amended technique must be used.  The best definition would simultaneously divide quasars spectroscopically and physically.  However, if additional spectral properties cannot break the degeneracy in the lower-left of Fig. \ref{fig:linecomp}, it may even be possible that multiple categorizations are required.

The authors would like to thank Steve Balbus, Tim Brandt, Forrest Collman, Martin Elvis, Jeremy Goodman, Daryl Haggard, Julian Krolik, Greg Novak, Jerry Ostriker, Guido Risaliti, Malte Schramm, Ohad Shemmer, Yue Shen, David Spergel, Michael Strauss, and Todd Thompson for valuable comments.  This work was supported by World Premier International Research Center Initiative (WPI Initiative), MEXT, Japan.


\end{document}